# CoverBLIP: scalable iterative matched filtering for MR Fingerprint recovery

Mohammad Golbabaee[1], Zhouye Chen[2], Yves Wiaux[2], and Michael Davies[1]

[1]School of Engineering, Institute for Digital Communications, University of Edinburgh, [2]School of Engineering and Physical Sciences, Institute for Sensors Signals and Systems, Heriot-Watt University

**Synopsis:**

Current popular methods for Magnetic Resonance Fingerprint (MRF) recovery are bottlenecked by the heavy computations of a matched-filtering step due to the size and complexity of the fingerprints dictionary. In this abstract we investigate and evaluate the advantages of incorporating an accelerated and scalable Approximate Nearest Neighbour Search (ANNS) scheme based on the Cover trees structure to shortcut the computations of this step within an iterative recovery algorithm and to obtain a good compromise between the computational cost and reconstruction accuracy of the MRF problem.

**Purpose:**

Current proposed solutions for the high dimensionality of the MRF reconstruction problem[1] rely on a linear compression step to reduce the matching computations[2,3] and boost the efficiency of fast but non-scalable searching schemes such as the KD-trees[4]. However such methodologies often introduce an unfavourable compromise in the estimation accuracy when applied to nonlinear data structures such as the manifold of Bloch responses with possible increased dynamic complexity and growth in data population[5].

To address this shortcoming we propose an *inexact* iterative reconstruction method, dubbed as the Cover BLoch response Iterative Projection (CoverBLIP). Iterative methods improve the accuracy of their non-iterative counterparts[6] and are additionally robust against certain accelerated approximate updates, *without* compromising their final accuracy[7,8]. Leveraging on these results, we accelerate matched-filtering using an ANNS algorithm based on Cover trees[9] with a robustness feature against *the curse of dimensionality*.

**Algorithm:**
The CoverBLIP iterations consist of
$$X^{t+1} = \widetilde{\mathcal{P}_D}\big(X^t - \mu A^H(A(X^t) - Y)\big),$$
where, $Y \in \mathbb{C}^{m \times L}$ is the undersampled k-space measurements across $L$ temporal frames, $\mu$ is the step-size (with line-search[6]), $X \in \mathbb{C}^{n \times L}$ is the spatio-temporal images reconstructed at iteration t, and $D \in \mathbb{C}^{d \times L}$ denotes the dictionary of $d \gg L$ fingerprints. The forward-adjoint operators $A, A^H$ correspond to the gradient updates and model the multi-coil sensitivities and the per-frame subsampled 2D Fourier Transforms. $\widetilde{\mathcal{P}_D}(.)$ is the *approximate* matched-filtering similar to[4,6] consisting of i) a search over the normalized dictionary $\overline{D}$ to replace temporal pixels of $X^{t+1}$ with their (approximate) nearest fingerprints, and ii) a proton density rescaling. For the search step we however use Cover trees' fast $(1+\epsilon)$-ANNS algorithm with controlled approximation levels $\epsilon \geq 0$[8,9]. For smooth O(1)-dimensional manifold data, Cover tree's complexities namely the storage, (offline) construction and ANNS times scale as $O(d), O(d \log(d))$ and $O(\epsilon^{-1} \log(d))$, respectively[9,10]. Remarkably the search complexity grows *logarithmically* with dataset population as compared to the linear complexity of a brute-force search used in [6].

When applicable and with a compromise in the accuracy, a temporal compression similar to[4] can be optionally used to shrink dimensions of $X^t, D$ across the $k \leq L$ dominant Eigen-components of $DD^H$. This also includes a compromise between cheaper distance evaluations during the search steps and the additional cost of iterative compression-basis applications (particularly when k has to be large and the gradient updates are using cheap FFT operations e.g. in Cartesian sampling).

**Methods:**

Methods are tested on a numerical brain phantom[11], a physical tube phantom and a healthy human brain (scanner data collected with a 12-channel head coil 3T GE MR750w scanner, GE Medical Systems, Milwaukee, WI). Sampling follows the multi-shot EPI-MRF protocol[6] (in contrast with the existing single-shot approaches[12,13]) with k-space compression factors x32 and x16 for the numerical and scanner data, respectively. The Numerical phantom is synthesized (256x256 sized T1, T2 maps in Figure 1 and off-resonance B0=0) with an Inversion Recovery (IR) Balanced SSFP sequence with L=1000 Flip Angles (FA), TR and TE similar to[3]. The constructed dictionary consists of 321'640 fingerprints for combinations of T1=[100:40:2000,2200:200:6000], T2 = [20:2:100,104:4:200,220:20:600], B0=[-250:40:-190,-50:2:50,190:40:250]. The scanner data uses IR Quantitative Transient-state Imaging (QTI) sequence (TR=16 ms, Tinv = 18 ms, 22:5x22:5 cm FOV, 128x128 matrix size, 1.3 mm in-plane resolution and 5 mm slice thickness) with a linear ramp FA variation 1°-70° across L=500 frames[14]. Dictionary construction uses the EPG model[15] with 23'866 fingerprints for combined T1=[10:20:1900,2100:100:6000], T2=[20:1:120,122:2:200,210:10:600], where T1>T2.

**Results and discussion:**

We compare the reconstruction times and accuracies of three iterative methods: BLIP[6] with exact NNS using MATLAB's matrix product, KDBLIP with randomize KD-trees ANNS (approximations controlled by the number of Checks) using the FLANN package[16], and CoverBLIP using Cover trees' $(1+\epsilon)$-ANNS with a parallel MATLAB interface to[17]. Experiments are conducted on a desktop with 8 CPU-Cores and 64GB RAM.

Cover tree construction takes around 10 and 0.8 seconds for the IR-BSSFP and EPG full-size dictionaries. Results tested on the numerical phantom (Figures 2 and 3) show that in the full-size ambient dimension CoverBLIP significantly outperforms the two other methods: better time-accuracy than KDBLIP, and a similar accuracy to BLIP however with $O(10^4)$ cheaper operations for matched-filtering. Since the numerical phantom is low-rank, we also apply temporal compression (k=20) i.e. a numerically beneficial setup to the KD-trees: CoverBLIP and KDBLIP report comparable time-accuracy with acceleration factors x50 compared to BLIP. Results tested on the scanner data (Figures 4 and 5) and using a smaller EPG dictionary show comparable recovered maps for all methods. Here the computational gain of approximate methods is less visible due to using a smaller size EPG dictionary. Nonetheless, CoverBLIP still outperforms other methods in the total runtime with matched-filtering accelerations x2 and x5-7 compared to KDBLIP and BLIP, respectively.

**Acknowledgement:**

We thank Arnold Benjamin (Centre for Clinical Brain Sciences, University of Edinburgh) and Dr. Pedro Gomez (GE Global Research, Munich) for providing the multi-shot EPI-MRF scanner data. This work is partly funded by the EPSRC grant EP/M019802/1 and the ERC C-SENSE project (ERCADG-2015-694888).

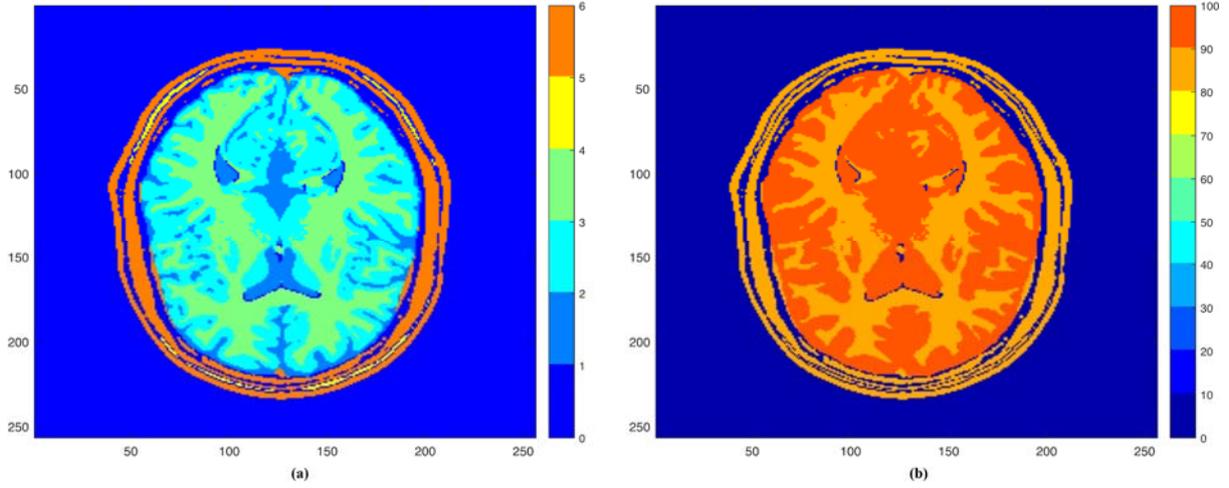

Figure 1: (a) the segmented anatomical brain phantom colored by index: 0 = background, 1 = CSF（T1=5012ms, T2=512ms）, 2 = grey matter（T1=1545ms, T2=83ms）, 3 = white matter（T1=811ms, T2=77ms）, 4 = adipose （T1=530ms, T2=77ms）, 5/6 = skin/muscle（T1=1425ms, T2=41ms）, (b) the proton density map used for generating the low-rank numerical phantom.

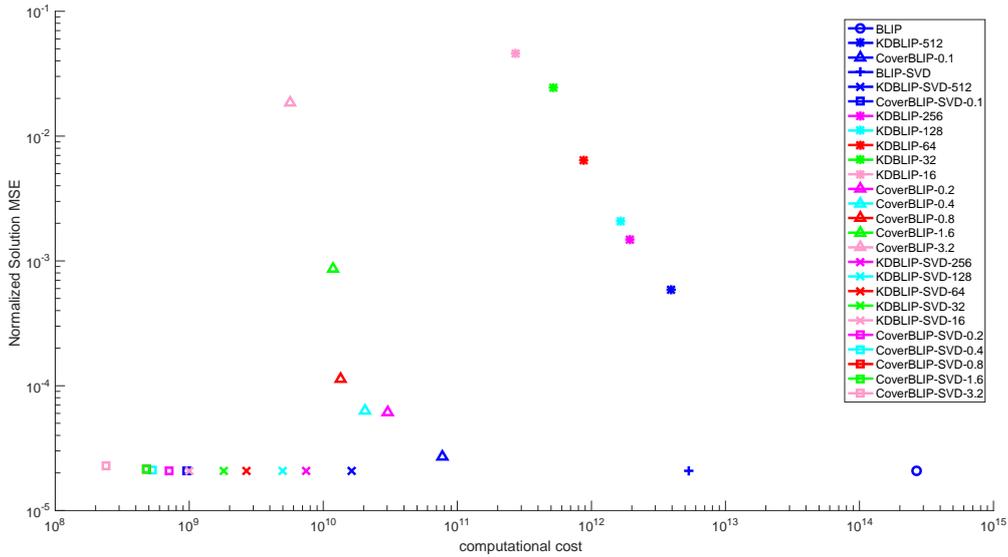

Figure 2: Computational cost (matched-filtering only) vs. normalized solution MSE (i.e. $\frac{\|\hat{X}-X\|}{\|X\|}$) for BLIP, KDBLIP and CoverBLIP (different approximation levels for CoverBLIP and KDBLIP), tested on the numerical phantom. The cost of matched-filtering is measured by the total number of the pairwise distance calculations (until convergence) times their dimensions. Using temporal compression (k=20), CoverBLIP performs comparable (or better) than KDBLIP. In full-size ambient dimension (L=1000) however, the computational cost of CoverBLIP is about $O(10^2)$ and $O(10^4)$ less than KDBLIP and BLIP, respectively.

|           | checks/eps | NMSE     | T1 error (ms) | T2 error (ms) | computational cost | Recovery time (s) |
|-----------|------------|----------|---------------|---------------|--------------------|-------------------|
| without SVD |          |          |               |               |                    |                   |
| BLIP      | -          | 2.09E-05 | 2.78          | 0.73          | 2.68E+14           | 8089.57           |
| KDBLIP    | 512        | 5.84E-04 | 11.23         | 1.37          | 3.94E+12           | 420.70            |
| CoverBLIP | 0.2        | 6.07E-05 | 7.58          | 0.93          | 3.02E+10           | 100.35            |
| SVD, k=20 |            |          |               |               |                    |                   |
| BLIP      | -          | 2.09E-05 | 2.78          | 0.73          | 5.35E+12           | 661.90            |
| KDBLIP    | 8          | 2.08E-05 | 2.78          | 0.73          | 8.87E+08           | 13.98             |
| CoverBLIP | 0.2        | 2.08E-05 | 2.78          | 0.73          | 7.03E+08           | 13.97             |

Figure 3: Statistical comparisons including the normalized solution MSE, average T1 and T2 errors (i.e. $\frac{1}{J}\Sigma_{j=1}^{J}|\widehat{T}_1(j) - T_1(j)|$), computational cost (matched-filtering only) and the total run times for BLIP, KDBLIP and CoverBLIP, tested on the numerical phantom (approximation levels with best runtime-accuracy are presented for CoverBLIP and KDBLIP). Regardless of the non-optimized code for cover tree ANNS, CoverBLIP and KDBLIP perform comparably when using temporal compression (k=20) i.e. a beneficial setup for the KD-trees. In full-size ambient dimension (L=1000) however, CoverBLIP outperforms all methods in recovery time with accelerations x4 and x80 compared to KDBLIP and BLIP, respectively.

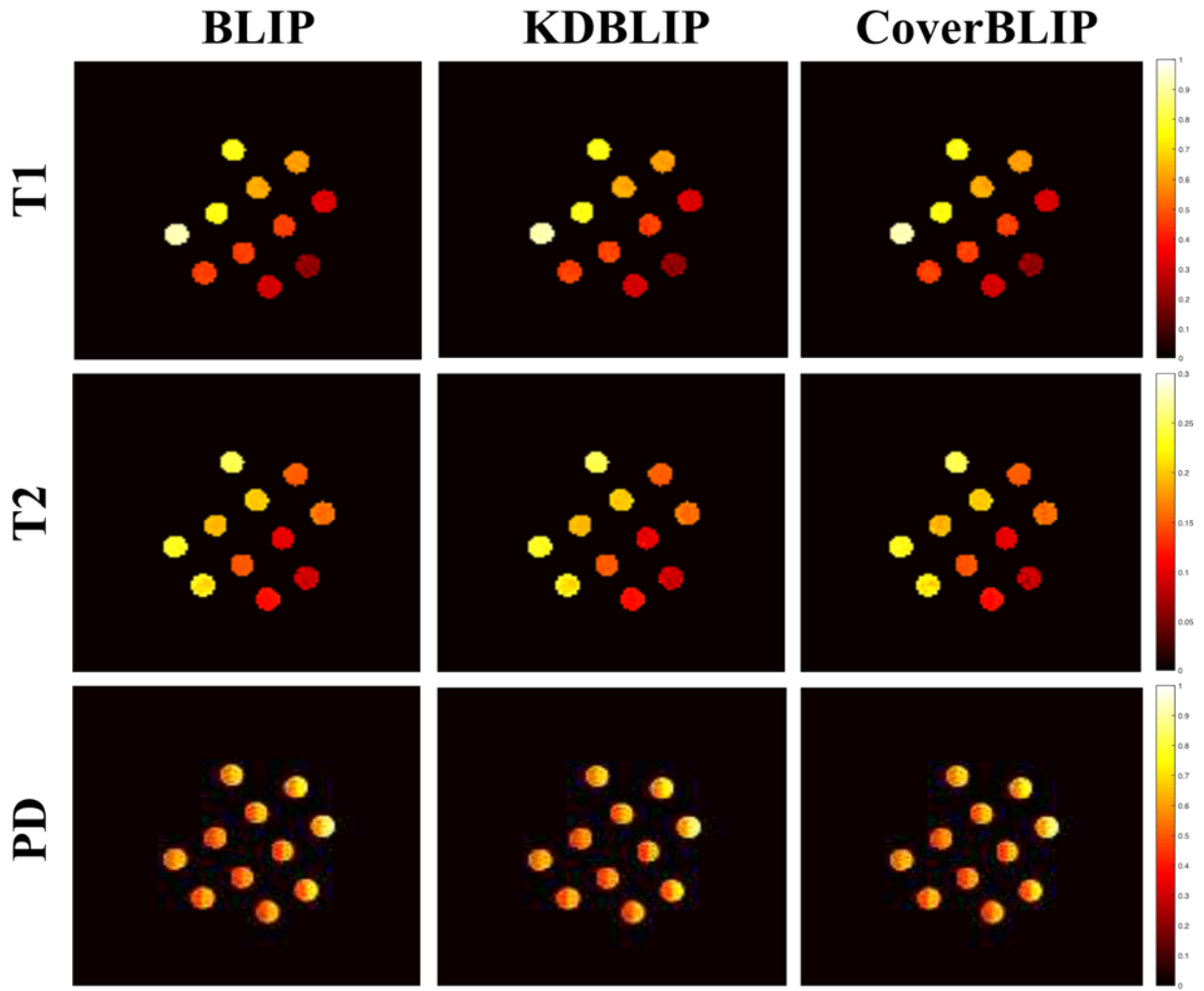

Figure 4: Recovered T1, T2 and PD maps for the tube phantom (Diagnostic Sonar, Livingston, UK) using BLIP, KDBLIP and CoverBLIP algorithms with no temporal compression. The iterative mathed-filtering steps take 20.0 seconds in BLIP, 5.3 seconds in KDBLIP (checks=64) and 2.8 seconds in CoverBLIP (epsilon=0.1).

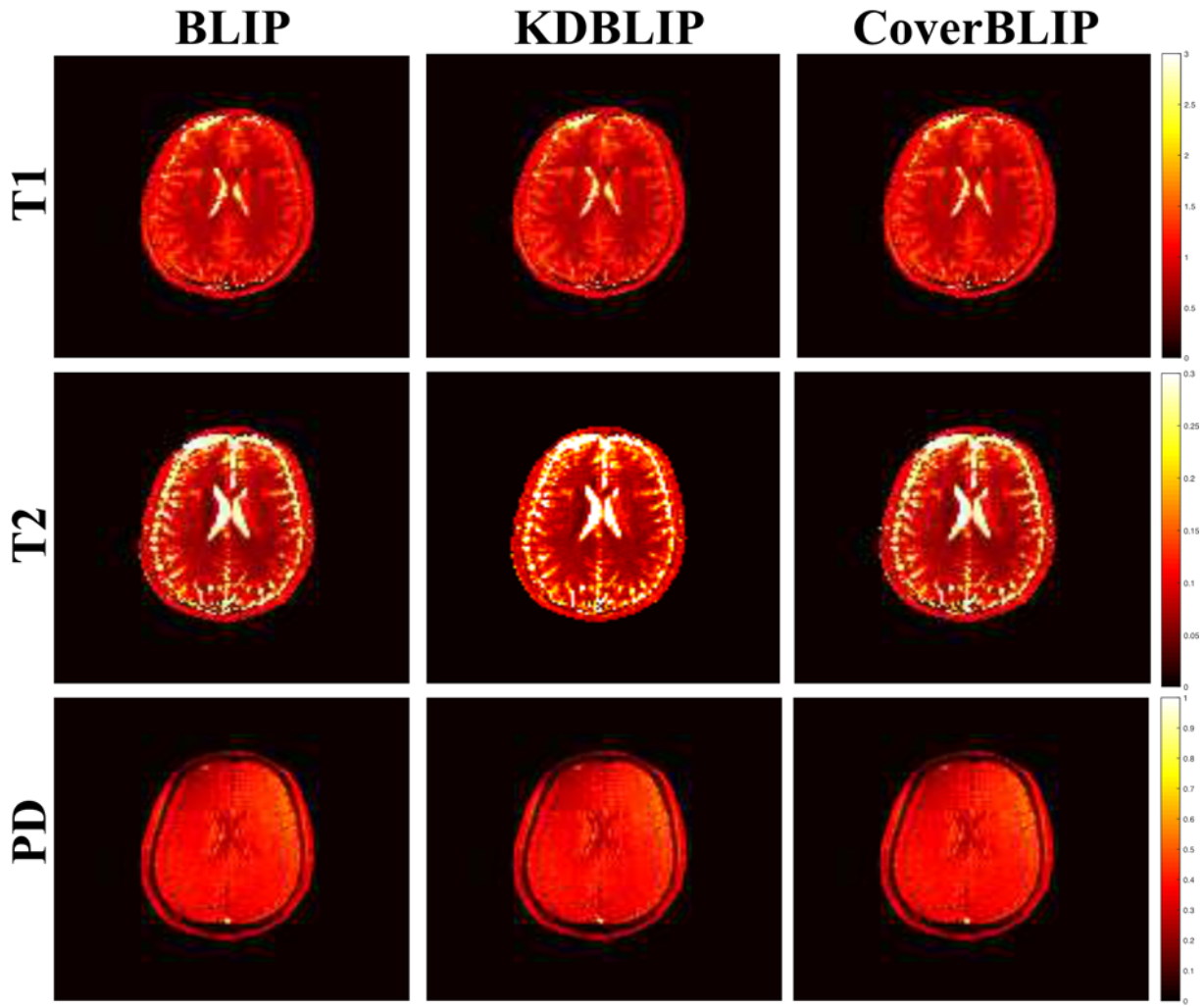

Figure 5: Recovered T1, T2 and PD maps for the human volunteer data using BLIP, KDBLIP and CoverBLIP algorithms with no temporal compression. The iterative mathed-filtering steps take 21.2 seconds in BLIP, 8.0 seconds in KDBLIP (checks=64) and 4.2 seconds in CoverBLIP (epsilon=0.1).